# Data-driven dissipative verification of LTI systems: multiple shots of data, QDF supply-rate and application to a planar manipulator

Tábitha E. Rosa and Bayu Jayawardhana

**Abstract** We present a data-driven dissipative verification method for LTI systems based on using multiple input-output data. We assume that the supply-rate functions have a quadratic difference form corresponding to the general dissipativity notion known in the behavioural framework. We validate our approach in a practical example using a two-degree-of-freedom planar manipulator from Quanser, with which we demonstrate the applicability of multiple datasets over one-shot of data recently proposed in the literature.

## 1 INTRODUCTION

In the design and control of complex high-tech systems, the provision of accurate models of the systems is often a time-consuming and costly process. Correspondingly, the data-driven approach has recently gained attention to complement the model-based methods.

In recent years, several approaches have been presented for analyzing systems properties directly from one-shot of input-output data that is sufficiently rich in information, or typically referred to as the persistently exciting property [1, 2, 3, 4]. The systems properties that can be verified directly from data include dissipativity [2, 5, 6, 7], the persistence of excitation properties [3], and data informativity [8]. In these cases, it is of utmost importance to have enough and suitable data to verify the information we desire. However, obtaining a single trajectory that provides such information on the system can be problematic in many real-life applications. In many cases, performing new or several experiments to get the proper data shot can be costly and time-consuming. In other cases, we can obtain missing or corrupted data in one-shot of data or get one-shot of poorly exciting data.

Tábitha E. Rosa, Bayu Jayawardhana
Engineering and Technology institute Groningen, University of Groningen, Nijenborgh 4, 9747 AG Groningen, The Netherlands e-mail: {`t.esteves.rosa,b.jayawardhana`}`@rug.nl`

Rather than obtaining systems properties directly from the data, one can first identify linear time-invariant (LTI) models before analyzing the systems. In this context, several techniques in systems identification that can deal with missing data or multiple datasets have been proposed, among many others, in [9, 10]. These approaches have been applied, for instance, in [11, 12].

Inspired by these results from systems identification literature, there are recent results on data-driven analysis using multiple-shots of input-output data, see for instance [3, 13, 14]. In [14], the authors propose, among others, a method to verify the persistence of excitation of a trajectory that recovers a set of multiple trajectories aligned over at least a certain number of steps. In [3], the datasets can come from different batches of measurements or experiments without the need to overlap.

As our main contribution, we extend the data-driven dissipative analysis using one-shot data to multiple shots of data following the approach pursued in [3]. In particular, we generalize the data-driven verification of dissipativity with quadratic input-output supply-rate function as presented in [1, 2, 5], which is extended recently in [5] to the quadratic differential form (QDF) of supply-rate function. The use of QDFs generalizes the well-known concept of QSR dissipativity and broadens the applicability of the dissipativity notion beyond passive systems and $\ell_2$ stable systems. For instance, it encompasses negative-imaginary systems [15, 16] and clockwise systems [17, 18]. Apart from the study and control of this class of systems, QDFs have also found application, for instance, in the stability analysis of mass-spring-damper systems with hysteresis [17] and in fault detection systems [19]. For the rest of this work, we will consider the time-difference form of the QDF.

In recent data-driven verification literature, it is common to consider a finite-time horizon in the summation of the supply-rate function, which can be verified easier than the infinite horizon case employing Willems' fundamental lemma. In this context, when the time horizon of dissipativity is limited to an integer $L > 0$, a data-driven verification method of *L-QSR dissipativity* is proposed in [2] using only one-shot of data which is convex in its formulation and whose data are required to be persistently excited. This method resolves the non-convexity issue in [1], which proposes a method that also incorporates the time-difference form in the QDF. When the largest time-difference in the QDF is given by an integer $N > 0$, the authors in [1] study the data-driven verification of the so-called $(L, N)$-*QSR dissipativity* via the behavioural framework. One crucial aspect in the verification of $L$-QSR dissipativity in [2] is that the condition must be checked not only for the time horizon $L$ but also for the time horizon $L - \nu$ for all admissible $\nu$, which we refer to in the work as the $(L, \nu)$-*QSR dissipativity*. Motivated by these results, the concept of $(L, \nu)$-QSR and $L$-QSR dissipativity are extended to the time-difference formulation of QDF in [5] using the notion of $(L, \nu, N)$-QSR and $(L, N)$-QSR dissipativity. In Section II, we present a precise definition of these notions. In these results, the need to test all admissible $\nu$ and the extension to the infinite-time horizon case is not trivial.

Correspondingly, the contributions of this work are as follows. Firstly, we propose a data-driven verification method employing multiple datasets whose combination is collectively persistently exciting. Secondly, we consider the *QSR*-dissipativity using the time-difference form of the supply-rate. Thirdly, using a single computable criterion, the proposed method can verify $(L, N)$-QSR dissipativity without the need for verification at different time horizons $L - \nu$ for all admissible $\nu$.

*Notation:* The set of vectors (matrices) of order $n$ ($n \times m$) with real entries is represented by $\mathbb{R}^n$ ($\mathbb{R}^{n \times m}$) and correspondingly, that with integer entries is denoted by $\mathbb{Z}^n$ ($\mathbb{Z}^{n \times m}$). Similar notation is applied to denote a vector (matrix) with zeros and ones by $0^n$ and $1^n$ (or $0^{n \times m}$ and $1^{n \times m}$), respectively. The $n \times n$ identity matrix is denoted by $I^n$. The set of positive (or negative) integers is denoted by $\mathbb{Z}_+$ (or $\mathbb{Z}_-$, respectively). The Kronecker product is indicated by $\otimes$. The space of square-summable discrete-time signals is denoted by $\ell_2(\mathbb{R}^\bullet)$. Given $e \in \ell_2(\mathbb{R}^\bullet)$, we denote

$$e_{[i,j]} = \begin{bmatrix} e(i)^\top & e(i+1)^\top & \cdots & e(j)^\top \end{bmatrix}^\top. \tag{1}$$

For a set of $q$ vectors $e^i_{[0,T_i-1]}, i = 1, \ldots, q$, we define

$$\mathbf{e}_{[0,\mathbf{T}-1]} := \begin{bmatrix} {e^1_{[0,T_1-1]}}^\top & {e^2_{[0,T_2-1]}}^\top & \cdots & {e^q_{[0,T_q-1]}}^\top \end{bmatrix}. \tag{2}$$

A Hankel matrix with $L \in \mathbb{Z}_+$ block rows of a finite sequence $e_{[0,T-1]}$ is given by

$$H_L(e_{[0,T-1]}) = \begin{bmatrix} e(0) & e(1) & \cdots & e(T-L) \\ e(1) & e(2) & \cdots & e(T-L+1) \\ \vdots & \vdots & \ddots & \vdots \\ e(L-1) & e(L-2) & \cdots & e(T-1) \end{bmatrix}. \tag{3}$$

**Lemma 1** *Finsler's Lemma [20] If there exist $w \in \mathbb{R}^n$, $Q \in \mathbb{R}^{n \times n}$, $B \in \mathbb{R}^{m \times n}$ with $\mathrm{rank}(B) < n$ and $B^\perp$ is a basis for the null space of $B$, that is, $BB^\perp = 0$, then all the following conditions are equivalent: (i). $w^\top Q w < 0$, $\forall w \neq 0 : Bw = 0$; (ii). ${B^\perp}^\top Q B^\perp < 0$; (iii). $\exists \mu \in \mathbb{R} : Q - \mu B^\top B < 0$; and (iv). $\exists \mathcal{X} \in \mathbb{R}^{n \times m} : Q + \mathcal{X} B + B^\top \mathcal{X}^\top < 0$.*

## 2 Problem Formulation

We consider the following discrete-time linear time-invariant (LTI) system

$$\Sigma : \begin{cases} x(k+1) = Ax(k) + Bu(k), & x(0) = x_0, \\ y(k) = Cx(k) + Du(k), \end{cases} \tag{4}$$

where $x(k) \in \mathbb{R}^n$ corresponds to the state vector, $u(k) \in \mathbb{R}^m$ is the control input and $y(k) \in \mathbb{R}^p$ is the output, with $u \in \ell_2(\mathbb{R}^m)$. We also assume that the state-space matrices are the minimal realization of the system $\Sigma$. We consider that we have access to the inputs and outputs for all instants of time $k = 0, \ldots, T_f$, where $T_f$ is any arbitrary given time.

In this work, we address the problem of verifying the $(L, N)$-dissipativity properties of $\Sigma$ following [1, 5]. For a given $\nu \geq n$, the system $\Sigma$ in (4) is said to be *(L, ν, N)-dissipative* with respect to a supply-rate $w(y_{[k,k+N]}, u_{[k,k+N]})$ if $\sum_{k=0}^{L-\nu-1} w(u_{[k,k+N]}, y_{[k,k+N]}) \geq 0$ holds for all trajectories $(u_{[0,L+N-1]}, y_{[0,L+N-1]})$ with $x_0 = 0$. Furthermore, it is called *(L, N)-dissipative* if $\Sigma$ is *(L, ν, N)*-dissipative for all $n \leq \nu < L$. Following the QDF formulation in [21, 5], we consider a quadratic difference form of supply function $w$ as follows

$$w(y_{[k,k+N]}, u_{[k,k+N]}) = \sum_{i,j=0}^{N} \begin{bmatrix} y(k+i) \\ u(k+i) \end{bmatrix}^\top \Phi_{ij} \begin{bmatrix} y(k+j) \\ u(k+j) \end{bmatrix} \quad (5)$$

for every $k \geq 0$, where each $\Phi_{ij}$ is the usual *QSR* matrix given by $\begin{bmatrix} Q_{ij} & S_{ij} \\ S_{ij}^\top & R_{ij} \end{bmatrix}$ with $Q_{ij} = Q_{ij}^\top$ and $R_{ij} = R_{ij}^\top$. Note that the $N$ in $(L, N)$-QSR dissipativity refers to the largest time differences in the QDF of the supply function. The matrix $\Phi_{ij}$ above gives the dissipativity relationship between a pair of data $(y(k+i), u(k+i))$ and $(y(k+j), u(k+j))$ for a given time shift $i$ and $j$. For passive and $\ell_2$ systems, they satisfy the supply-rate (5) with $N = 0$. We note here that in the formulation of quadratic programming, later on, we use the combination of all $\Phi_{ij}$ via $\Phi_N = \begin{bmatrix} \Phi_{00} & \cdots & \Phi_{0N} \\ \vdots & \ddots & \vdots \\ \Phi_{N0} & \cdots & \Phi_{NN} \end{bmatrix}$ where $\Phi_{ji} = \Phi_{ij}^\top$ for all $i, j = \{0, \ldots, N\}$. Specifically, for every $\nu$, we will refer the $(L, \nu, N)$ dissipative property with the supply function $w$ given by (5) as $(L, \nu, N)$-*QSR dissipativity*.

An important assumption used in the literature (as in [1, 2, 5]) for verifying *L-QSR-dissipativity* is the persistent excitation of the input data. The main idea of having the input measurements persistently exciting is that by using a single shot of data, we can obtain the rest of the admissible trajectories of (4), which is known as Willems' fundamental lemma. In practice, where missing data, corrupted data or an insufficient amount of data can take place, the available one-shot of data $z_{[0,T-1]}$ may not satisfy the central hypothesis in this lemma (namely, the persistence of excitation condition). Consequently, we cannot obtain information on the rest of the admissible trajectories. To deal with such cases, the following *collectively persistently exciting* notion is introduced in [3], which plays an essential role in the present work.

**Definition 1** *[3]* Consider a set of $q$ measured trajectories given by $\mathbf{e}_{[0,\mathbf{T}-1]}$ as in (2). This set of trajectories is *collectively persistently exciting* of order $L$ with $0 < L \leq T_i$ for all $i = 1, 2, \ldots, q$, if the following mosaic-Hankel matrix

$$\mathcal{H}_L(\mathbf{e}_{[0,\mathbf{T}-1]}) = \begin{bmatrix} H_L(e^1_{[0,T_1-1]}) & H_L(e^2_{[0,T_2-1]}) & \cdots & H_L(e^q_{[0,T_q-1]}) \end{bmatrix}, \quad (6)$$

where $\mathbf{e}_{[0,\mathbf{T}-1]}$ is the set of $q$ shots of trajectories, and has full row rank.

Using this notion, a set of measured trajectories with possible different lengths can be used together to obtain every possible trajectory of $\Sigma$ with a time horizon $L$, as given in the following lemma.

**Lemma 2** *[3]* Suppose that $(\mathbf{y}_{[0,\mathbf{T}-1]}, \mathbf{u}_{[0,\mathbf{T}-1]})$, for $q$ snapshots, are trajectories of $\Sigma$. If the set of inputs $u^i$ is collectively persistently exciting of order $L + n$ with $0 < L \leq T_i$ for all $i = 1, 2, \ldots, q$, then $(\bar{y}_{[0,L-1]}, \bar{u}_{[0,L-1]})$ is an admissible trajectory of (4) if and only if there exists a vector $\alpha \in \mathbb{R}^{q(1-L)+\sum_{i=1}^{q} T_i}$ such that

$$\begin{bmatrix} \mathcal{H}_L(\mathbf{y}_{[0,\mathbf{T}-1]}) \\ \mathcal{H}_L(\mathbf{u}_{[0,\mathbf{T}-1]}) \end{bmatrix} \alpha = \begin{bmatrix} \bar{y}_{[0,L-1]} \\ \bar{u}_{[0,L-1]} \end{bmatrix}. \quad (7)$$

Given the information mentioned above, in this work, we address the problem of verifying the dissipativity with the QDF supply-rate function of a system $\Sigma$ based on multiple shots of measured trajectories which are collectively persistently excited.

**Multiple shots $(L, \nu, N)$-QSR dissipativity verification problem:** Given $q$ multiple shots of trajectories $(\mathbf{y}_{[0,\mathbf{T}+N-1]} \mathbf{u}_{[0,\mathbf{T}+N-1]})$, of $\Sigma$, verify whether $\Sigma$ is $(L, \nu, N)$-QSR dissipative for some $\nu \geq n$ with $0 < L \leq T_i$ for all $i = 1, \ldots q$.

## 3 Main result

In the following theorem, we present our main result where a data-driven verification method is given to check the $(L, N)$-QSR-dissipativity of $\Sigma$. For any time $k \geq 0$,

$$Z(k) := \begin{bmatrix} y(k)^\top & u(k)^\top & \cdots & y(k+N)^\top & u(k+N)^\top \end{bmatrix}^\top. \quad (8)$$

Using this windowed data vector of size $N$, the $i$-th batch data $Z^i_{[0,T_i-1]}$ is understood as in (1) and the expression of Hankel matrix $H_L(Z^i_{[0,T_i-1]})$ follows the composition in (3). Following the expression of mosaic-Hankel matrix in Definition 1, we define the mosaic-Hankel matrix $\mathcal{H}_L(\mathbf{Z}_{[0,\mathbf{T}-1]})$ as in (6) using $q$ snapshots of data $\mathbf{Z}_{[0,\mathbf{T}-1]}$, following . Additionally, we denote

$$U = \begin{bmatrix} U_{\text{aux}} & 0^{(m+p)\nu \times (m+p)(L-\nu)(N+1)} \end{bmatrix}, \quad U_{\text{aux}} = I^\nu \otimes \begin{pmatrix} I^{m+p} & 0^{(m+p) \times (m+p)N} \end{pmatrix}. \quad (9)$$

**Theorem 1** *Let the integer $L > 0$ and $(\mathbf{y}_{[0,\mathbf{T}-1]}, \mathbf{u}_{[0,\mathbf{T}-1]})$, be a set of $q$ trajectories of (4) with n be the order of the system, $\sum_{i=1}^{q} T_i \geq (L-1)(m+p+q)$ and $n+1 < L \leq T_i$ for all $i = 1, \ldots, q$. Suppose that the set of $q$ snapshots of inputs $\mathbf{u}_{[0,\mathbf{T}+N-1]}$, is collectively persistently exciting of order $L + N + n$ and there exists $\nu$ s.t. $n \leq \nu < L$ and*

$$U_\perp^\top \mathcal{H}_L(\mathbf{Z}_{[0,\mathbf{T}-1]})^\top \Phi_L \mathcal{H}_L(\mathbf{Z}_{[0,\mathbf{T}-1]}) U_\perp \geq 0, \quad (10)$$

*holds, where $\Phi_L = I_L \otimes \Phi_N$, and $U_\perp = (U\mathcal{H}_L(\mathbf{Z}_{[0,\mathbf{T}-1]}))^\perp$ with $U$ given in (9). Then (4) is $(L, \lambda, N)$-QSR dissipative for all $\nu \leq \lambda < L$.*

***Proof*** First, by the hypotheses of the theorem, we have that the set of $q$ snapshots of inputs $\mathbf{u}_{[0,\mathbf{T}+N-1]}$, is collectively persistently exciting of order $L + N + n$ holds with $L + N \leq T_i$, for all $i = 1, \ldots, q$, such that (10) holds for an integer $\nu$ in the interval $[n, L)$. Following the main results in [3], it follows from Definition 1 and Lemma 2 that for a given set of $q$ trajectories $(\mathbf{y}_{[0,\mathbf{T}+N-1]}, \mathbf{u}_{[0,\mathbf{T}+N-1]})$, if $u$ is collectively persistently exciting then other admissible trajectory $(\bar{y}_{[0,L+N-1]}, \bar{u}_{[0,L+N-1]})$ of $\Sigma$

satisfies

$$\begin{pmatrix} \mathcal{H}_{L+N}\left(\mathbf{y}_{[0,\mathbf{T}+N-1]}\right) \\ \mathcal{H}_{L+N}\left(\mathbf{u}_{[0,\mathbf{T}+N-1]}\right) \end{pmatrix} \alpha = \begin{bmatrix} \bar{y}_{[0,L+N-1]} \\ \bar{u}_{[0,L+N-1]} \end{bmatrix} \tag{11}$$

for some $\alpha$. Likewise, if (11) holds then we can rewrite it using $\mathbf{Z}$ such that

$$\mathcal{H}_L(\mathbf{Z}_{[0,\mathbf{T}-1]})\alpha = \bar{Z}_{[0,L-1]} \tag{12}$$

holds with the same $\alpha$ as in (11) and with $\bar{Z}_{[0,L-1]}$ be constructed through the rearrangement and stacking of the elements in $\left(\bar{y}_{[0,L+N-1]}, \bar{u}_{[0,L+N-1]}\right)$ as used in $\mathbf{Z}_{[0,\mathbf{T}+N-1]}$. In other words, verifying (12) is the equivalent of verifying (11). Thus, the matrices multiplying $\alpha$ in (11) and (12) share the same rank, meaning that we can verify if the persistence of excitation conditions hold using any of these matrices.

From the hypotheses of the theorem, we have that $U_\perp$ is a null space of $U\mathcal{H}_L(\mathbf{Z}_{[0,\mathbf{T}-1]}))$, e.g., $U\mathcal{H}_L(\mathbf{Z}_{[0,\mathbf{T}-1]})U_\perp = 0$ holds. Combining this fact with (12) implies that $U\mathcal{H}_L(\mathbf{Z}_{[0,\mathbf{T}-1]})\alpha = 0$, for any $\alpha$ that satisfies (12) with zero initial condition $[\bar{y}_{[0,\nu-1]}^\top \bar{u}_{[0,\nu-1]}^\top]^\top = 0$. Note that this happens due to the definition of $U$ as in (9), as the non-zero elements are in place to ensure that the initial conditions are null. Using conditions 1) and 2) from Finsler's lemma in Lemma 1, and having in mind the aspects above of $U_\perp$, the inequality in (10) is equivalent to $\alpha^\top \mathcal{H}_L(\mathbf{Z}_{[0,\mathbf{T}-1]})^\top \Phi_L \mathcal{H}_L(\mathbf{Z}_{[0,\mathbf{T}-1]})\alpha \geq 0$, for all $\alpha$ as before (which results in admissible trajectories with zero initial conditions). It follows immediately that $\sum_{k=0}^{L-1} \bar{Z}(k)^\top \Phi_N \bar{Z}(k) = \sum_{k=0}^{L-1} \sum_{i,j=0}^{N} \begin{bmatrix} \bar{y}(k+i) \\ \bar{u}(k+i) \end{bmatrix}^\top \Phi_{ij} \begin{bmatrix} \bar{y}(k+j) \\ \bar{u}(k+j) \end{bmatrix} \geq 0$, which is the equivalent of verifying $\sum_{k=0}^{L-1} \bar{Z}(k)^\top \Phi_N \bar{Z}(k) = \bar{Z}_{[0,L-1]}^\top \Phi_L \bar{Z}_{[0,L-1]} \geq 0$ for all trajectories $\left(\bar{y}_{[0,L+N-1]}, \bar{u}_{[0,L+N-1]}\right)$ with initial conditions $[\bar{y}_{[0,\nu-1]}^\top \bar{u}_{[0,\nu-1]}^\top]^\top = 0$. Consequently, considering the $\nu$ initial conditions, the previous results are equivalent to verifying whether

$$\sum_{k=0}^{L-\nu-1} \tilde{Z}(k)^\top \Phi_N \tilde{Z}(k) = \sum_{k=0}^{L-1} \bar{Z}(k)^\top \Phi_N \bar{Z}(k) \geq 0 \tag{13}$$

holds for any trajectory $\left(\tilde{y}_{[0,L+N-\nu-1]}, \tilde{u}_{[0,L+N-\nu-1]}\right)$ with initial conditions $\tilde{x}_0 = 0$, where $x$ is the state of an arbitrary minimal realization of $\Sigma$, and for any trajectory $\left(\bar{y}_{[0,L+N-1]}, \bar{u}_{[0,L+N-1]}\right)$ with initial conditions $[\bar{y}_{[0,\nu-1]}^\top \bar{u}_{[0,\nu-1]}^\top]^\top = 0$. Additionally, we have that if (13) holds for some value of $\nu$, than we recover the definition of $(L, \nu, N)$-$QSR$-dissipativity, as presented in [2] and [5].

Now it remains to prove that $(L, \lambda, N)$-$QSR$ dissipativity holds for all $\lambda > \nu$. For this purpose, let us take $\nu < \lambda < L$. We will now show that the solvability of (10) also implies that we have $(L, \lambda, N)$-$QSR$ dissipativity, i.e., (10) holds with $U$ and $U_{\text{aux}}$ defined in (9) to be replaced by $U_\lambda = \begin{bmatrix} U_{\text{aux}} & 0^{(m+p)\lambda \times (m+p)(L-\lambda)(N+1)} \end{bmatrix}$, $U_{\text{aux},\lambda} = I^\lambda \otimes \left(I^{m+p} \ 0^{(m+p)\times (m+p)N}\right)$, respectively. First we note that the null space $U_{\lambda,\perp} := (U_\lambda \mathcal{H}_L(\mathbf{Z}_{[0,\mathbf{T}-1]}))^\perp$ with the above choice of $U_\lambda$ exists if $\sum_{i=1}^{q} T_i \geq \lambda(m+p) + (L-1)q$. By the hypothesis of the theorem where we assume that $\sum_{i=1}^{q} T_i \geq (L-1)(m+p+q)$ and $\lambda < L$, it follows immediately that $\sum_{i=1}^{q} T_i \geq \lambda(m+p) + (L-1)q$ holds. Thus the null space $U_{\lambda,\perp}$ is non-empty. By the construction of $U$ in (9) and $U_\lambda$ with $\nu < \lambda$, it follows that $\text{Im}(U) \subset \text{Im}(U_\lambda)$. This implies that $U_{\lambda,\perp} \subset U_\perp$. Consequently, if (10) holds for some value of $\nu$ then (10) also holds with $U$ and $U_{\text{aux}}$ be replaced by $U_\lambda$ and $U_{\text{aux},\lambda}$ for all $\lambda > \nu$. This proves the claim. □

One direct implication from Theorem 1 is that when (10) holds with $\nu = n$ then we obtain the $(L, N)$-QSR dissipativity of $\Sigma$. Consequently, we can use $\nu$ to upper-bound the order of the system $n$ as discussed in [2, 5].

Theorem 1 covers existing results in [5, 2], which is restricted to the one-shot data case. On the one hand, by taking $q = 1$, we immediately recover the results in [5] with the main difference on the assumption of $T$. When we apply Theorem 1 in [5] using the hypotheses in Theorem 1 above, then we immediately verify the $(L, N)$-$QSR$ dissipativity of the system without the need to check for each $\nu$ as posited in [5]. On the other hand, by setting $N = 0$ and taking $q = 1$, we recover the $L$-$QSR$-dissipativity, which is also studied in [2]. The idea in Theorem 1 and that in [5] share many commonalities with the approach in [2]. Note, however, that the main difference is in how we construct the main inequality and arrange the data used in the numerical verification.

By modifying the matrices $Q_{ij}$, $S_{ij}$ and $R_{ij}$ in $\Phi_{ij}$, one can directly verify various systems properties of $\Sigma$ that involve particular dissipative inequality as discussed in the Introduction. This includes passive systems, $\ell_2$-stable systems, negative-imaginary/counter-clockwise systems, and positive-imaginary/clockwise systems. Additionally, we can potentially extend this result to the case where the system is affected by noise by applying, for instance, the results in [22].

## 4 Example - 2 degree-of-freedom planar manipulator

For validating our technique experimentally, we consider experiments performed in a two-degree-of-freedom (DoF) planar manipulator from Quanser [23], using a rigid joint configuration as detailed in [24]. As presented in [24], the robot is modelled by state equations with four state variables representing the generalized positions and momenta of the two joints, with input variables of electrical current (in Amperes) to the actuation motors and with output variables of the generalized position of the joints. From the property of Euler-Lagrange systems [25], it is well-known that the open-loop behaviour of any robotic manipulator without damping (lossless) is dissipative with respect to the QDF supply-rate function of $w(y(t), u(t)) = \dot{y}(t)^\top u(t)$ (in the continuous-time case) or its associated discrete-time version $w(y_{[k,k+N]}, u_{k,k+N}) = (y(k+1) - y(k))^\top u(k+1)$. In our data-driven dissipative verification, we consider a closed-loop system using the controller proposed in Case 2 of Section 5.1 from [24], while the robot is operated in the neighbourhood of the generalized positions $q^* = \begin{bmatrix} 0.6 & 0.8 \end{bmatrix}^\top$ and generalized momenta $p^* = \begin{bmatrix} 0 & 0 \end{bmatrix}^\top$ so that the dynamics can be approximated by an LTI system (4). In this case, the QDF supply-rate function of the closed-loop system will also contain dissipation terms introduced by the friction damping and feedback controller.

For generating the input-output data, we excite the system with an external signal with a normal distribution, a standard deviation of 0.05 and a zero-mean that is calculated and applied separately to each joint. The time-series input-output data is collected with a sampling time of $T_s = 0.005$s. Figure 1 shows the data from one of the experiments where we can see that the robots operate in the neighbourhood of $(q^*, p^*)$. Note that the input for the second motor is saturated. A video of this experiment can be found at youtu.be/2kg4Tp3qp3Y.

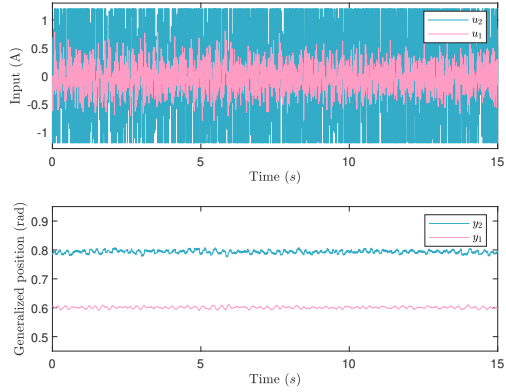

**Fig. 1** One set of experimental data of a 2-DoF planar manipulator from Quanser. The closed-loop system (robot + controller) operates at an equilibrium point $(q^*, p^*) = (\begin{bmatrix} 0.6 \\ 0.8 \end{bmatrix}, \begin{bmatrix} 0 \\ 0 \end{bmatrix})$ while perturbed by an external signal with a normal distribution, a standard deviation of 0.05 and zero-mean.

We consider five snapshots ($q = 5$) of data obtained from five different batches of experiments, representing the case where a batch of measurement data is unavailable. The size of the snapshots is given by $T_i \in \{10, 9, 13, 10, 12\}$, where each of them is taken from a different point of the experiment and each snapshot has the form $(\mathbf{u}_{[0, \mathbf{T}+N-1]}, \mathbf{y}_{[0, \mathbf{T}+N-1]})$. Additionally, we take $L = 4$ and $N = 1$, where the choice of $L = 4$ comes from the knowledge of the order of the system and $N = 1$ is from the QDF supply-rate function of open-loop robotic manipulators, as discussed before.

We use the numerical software Matlab (R2020a) in conjunction with the parser YALMIP [26] and the solver Mosek [27] for the optimization procedures of finding feasible solutions $\Phi > 0$ in Theorem 1 via linear matrix inequality (LMI) conditions. Using the aforementioned multiple datasets, the obtained matrix $\Phi_1$ is given by

$$\Phi_1 = \mathrm{diag}\left\{1, 1, 1, 1, 1.999, 2, \begin{bmatrix} 1.463 & -0.098 \\ -0.098 & 0.674 \end{bmatrix}\right\}. \tag{14}$$

For comparison purposes, we apply the methods in [2] and [5] to check whether they can find a supply function to which the system is dissipative. Both of these methods are used for verifying the dissipativity, however, they consider using one single data shot with the input persistently excited. Thus, we test each of the small snapshots separately, using $N = 1$ and $N = 0$ for the method in [5] and $N = 0$ for [2]. As expected, the searches using all the snapshots of data separately cannot provide results since each dataset individually is not persistently excited.

We can verify numerically whether the identified supply function (14) holds for all batches of data. Particularly, we evaluate whether $\Upsilon(T_f) := \sum_{k=0}^{T_f} w(u(k), y(k)) \geq 0$ holds for arbitrary $T_f \geq 0$ with the identified $\Phi_1$ as in (14). This is presented in

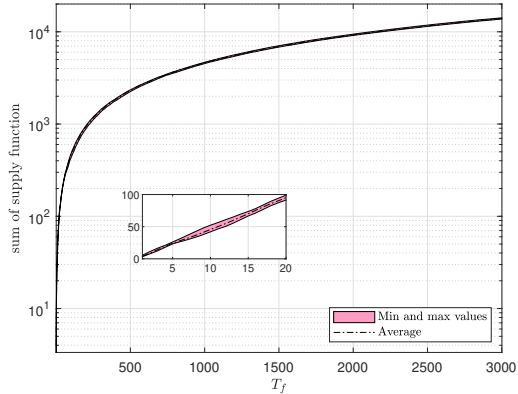

**Fig. 2** The plot of $\Upsilon(T_f) := \sum_{k=0}^{T_f} w(u(k), y(k))$ as a function of the time $T_f$, where the supply-rate function $w$ is given by (5) with the identified $\Phi_1$.

Figure 2, which shows numerically that the system is indeed dissipative with respect to $\Phi_1$ for an extended time horizon beyond the prescribed time horizon of $L$.

## 5 CONCLUSIONS

This work presents a data-driven method to verify dissipativity properties using a quadratic difference form of the supply-rate function and multiple data sets that collectively satisfy the persistent excitation condition. The method is validated on experimental data from a 2-DoF robotic manipulator. Future work includes a fault detection method using the concept of data-driven dissipativity analysis, in which we identify the dissipativity inequality using a data-driven approach similar to the one in this work.